\documentclass[12pt]{article}

\begin{document}

\title{
{\Large \bf Some boundary effects in quantum \\field theory
\footnote{To appear in Physical Review D 2004}} 
}         

\author{V. B. Bezerra \\
Universidade Federal da Paraiba, \\
Depto. de F\'\i sica \\
Caixa Postal 5008, 58051-970 - Jo\~{a}o Pessoa, PB, Brazil \\
e-mail: valdir@fisica.ufpb.br \\
and \\
M. A. Rego-Monteiro \\
Centro Brasileiro de Pesquisas F\'\i sicas, \\ 
Rua Xavier Sigaud 150, 22290-180 - Rio de Janeiro, RJ, Brazil\\
e-mail: regomont@cbpf.br 
}             
\date{} 
         
\maketitle

\begin{abstract}
\indent
We have constructed a quantum field theory in a finite box, 
with periodic 
boundary conditions, using the hypothesis that particles 
living in a finite box are created and/or annihilated by 
the creation and/or annihilation operators, respectively, 
of a quantum harmonic oscillator on a circle. An expression 
for the effective coupling constant is obtained showing 
explicitly its dependence on the dimension of the box.

\end{abstract}

\vspace{1cm}

\begin{tabbing}

\=xxxxxxxxxxxxxxxxxx\= \kill

field theory in a compact space; variation of the coupling 
constant. \\

\end{tabbing}

\newpage

\section{Introduction}

The fact that the energy eigenvalues of the quantum 
harmonic oscillator is given by $E_n = (n+1/2) \hbar w$
allow us to interpret their successive energy levels
as being obtained by the creation of a quantum particle of 
frequency $w$. This interpretation of the energy spectrum
of the quantum harmonic oscillator 
was successfully used in the second quantization
formalism. In short, one could say that Planck's 
hypothesis is realized in the second quantization 
formalism by the use of creation and annihilation
operators of the quantum harmonic oscillator system 
$^{\cite{tdlee}}$. This realization is obtained for
the quantum harmonic oscillator defined on an infinite line.

Let us consider a situation in which we want to describe 
the interaction of quantum particles living in a finite 
box with boundary conditions, for example, using the 
second quantization formalism.  In this context, it seems 
natural to assume the statement concerning the connection
between Planck's hypothesis and the energy levels of
a quantum harmonic oscillator in this finite space and
therefore analyze the consequences of this assumption 
in the construction of a quantum field theory (QFT) in a 
compact manifold. 

In \cite{circulo}  a discussion of a quantum
harmonic oscillator in a circle and its associated Heisenberg
algebra was presented. It was found that Mathieu's equation can 
satisfactorily describe the system and that the creation
and annihilation operators of the system satisfy a sort of
deformed Heisenberg algebra. In \cite{qft1} a construction of 
a deformed scalar QFT based on $q$-oscillator \cite{qosc}, 
which is a deformed Heisenberg algebra, was presented 
and in \cite{qft2} a procedure to perform perturbative computation up 
to second order in the coupling constant was implemented. 
Subsequently, it was shown \cite{qft3} that this deformed 
scalar quantum model is renormalizable up to second order 
in the coupling constant.

In this paper, we use the same procedure developed in Refs. 
\cite{qft1} and \cite{qft2} to perform a perturbative
computation for a QFT in a box. To do this, we
use the hypothesis, already mentioned, that in a 
compact space with periodic boundary conditions, particles 
are created and/or annihilated by the creation and/or 
annihilation operators of a Heisenberg algebra of the quantum
harmonic oscillator defined on a circle. As a result, 
we find that the effective coupling constant which appears 
in the perturbation series depends on a dimensionless quantity 
related to the linear 
dimension of the box. This approach permits us to construct a field 
theory that creates at any point of the space-time particles described
by a deformed Heisenberg algebra, which in the present case, the 
deformation parameter is inversely proportional to the dimension of 
the box. In this way we can investigate the interaction of point
particles in compact spaces, showing how the boundary affects this 
interaction.

Finally, we have computed the variation of the effective coupling 
constant for two different values of the size of the box, namely 
one corresponding to the time of nucleosynthesis of the standard 
cosmological model and the other to the present epoch. The choice of 
these values for the sizes of the box were done simply to perform a 
calculation and to show an example of the effect of the boundary
on the effective coupling constant. This does not mean 
that our model has a connection with the standard cosmological model.

This paper is organized as follows: In Section II, we present
a discussion of a quantum harmonic oscillator on a circle
which is described by Mathieu's equation. The deformed 
Heisenberg algebra associated with Mathieu's equation is 
presented in Section III. In Section IV, we present a 
construction of a QFT in a box and perform some perturbative
computation. In Section V we present the bound for the
variation of the coupling constant. Finally, in Section
VI we conclude with some comments.

\section{The quantum harmonic oscillator on a circle}

In this Section we are going 
to discuss an equation defined on a finite interval
of length $L$ which reproduces the ordinary quantum harmonic 
oscillator in the limit $L \rightarrow \infty$. 
For this proposal it is convenient to describe quantum 
mechanics on a periodic line and to do this we shall 
follow Ohnuki-Kitakado's formalism $^{\cite{ok1}}$. 
According to this formalism 
there are inequivalent quantum mechanics on $S^1$ (periodic
line) depending on a parameter $\alpha$ ($0 \leq 
\alpha < 1$). The momentum operator $G$ on $S^1$
in the coordinate representation is given in this
formalism as
$^{\cite{ok1}, \cite{ok2}}$
\begin{equation}
G \longrightarrow \frac{1}{i} \frac{d}{d\theta} +
\alpha \, , \,\,\,\,\,  0 \leq \alpha < 1 \, 
\label{eq:mom1}
\end{equation}
and the coordinate operator is given in terms of the
unitary operator $W$
\begin{equation}
W \longrightarrow e^{i \theta} \, .
\label{eq:coord1}
\end{equation}

Let us consider the following equation on $S^1$
\cite{circulo}:
\begin{equation}
G^2 \Psi + K \left[ W+ W^{\dagger} \right] \Psi 
= \epsilon \Psi \, ,
\label{eq:hos1}
\end{equation}
where $G$ and $W$ were already defined \footnote{
It would be also possible to define an equation with
quadratic powers of $W$ and $W^{\dagger}$, but the above
equation is the simplest one.}.
In order to have the above equation in the coordinate
representation
we substitute Eqs. (\ref{eq:mom1}) and (\ref{eq:coord1}) in
Eq. (\ref{eq:hos1}) for $\alpha = 0$. Thus we obtain
\begin{equation}
\frac{d^2 \Psi(\theta)}{d\theta^2} + (\epsilon - 2 K \cos\theta)
\Psi(\theta) = 0 \, ,
\label{eq:mathieu1}
\end{equation}
with $\Psi(\theta=0)=\Psi(\theta=2 \pi)$. 
This equation is the well known Mathieu's equation which first 
appeared in 1868
in the study of the vibrations of a stretched membrane of 
elliptic cross-section $^{\cite{mathieu1}}$. Mathieu's equation
is an important equation in physics arising from the study of
a variety of physical problems, from ordered crystals with the potential
$\cos 2x$ $^{\cite{mathieu2}}$ to the wave equation of scalar 
fields in the background of a D-brane metric $^{\cite{mathieu3}}$.
Note that this is one possible equation
on a periodic line since we chose for simplicity $\alpha =0$ in
Eq. (\ref{eq:mathieu1}). According to Ohnuki-Kitakado's formalism
$^{\cite{ok1}}$ there are inequivalent quantum mechanics on $S^1$ 
for each value of the parameter $\alpha$ ($0 \leq \alpha < 1$).

In order to consider the limit of Eq. (\ref{eq:mathieu1})
when the radius of the circle goes to infinity we perform the
change of variables
\begin{equation}
\theta = \frac{\pi}{L} y + \pi \, , \,\,\, -L \leq y \leq L \, .
\label{eq:mudvar}
\end{equation}
Using Eq. (\ref{eq:mudvar}), Eq. (\ref{eq:mathieu1}) becomes
\begin{equation}
\frac{d^2 \Psi}{d y^2} + \left(E + \frac{2\pi^2}{L^2} K 
\cos\frac{\pi}{L}y \right)
\Psi = 0 \, ,
\label{eq:mathieu3}
\end{equation}
where $E= \pi^2 \epsilon/L^2$. Then, using a trivial trigonometric
identity and calling $\lambda \equiv E+2 \pi^2 K/L^2$ we obtain
\begin{equation}
\frac{d^2 \Psi}{d y^2} + \left[ \lambda - \frac{\pi^4}{L^4} K y^2
\left( \frac{\sin\pi y/2L}{\pi y/2L} \right)^2 \right]
\Psi = 0 \, .
\label{eq:mathieu4}
\end{equation}
The well known Schr\"{o}dinger's equation for the 
quantum harmonic oscillator
\begin{equation}
\frac{d^2 \Psi}{d y^2} + \left( \lambda - y^2 \right)
\Psi = 0 \, ,
\label{eq:schroed}
\end{equation}
is obtained for $K=L^4/\pi^4$ in Eq. (\ref{eq:mathieu4}),
if $L \rightarrow \infty$. It is then reasonable to 
take Eq. (\ref{eq:mathieu1}) with $K=L^4/\pi^4$ as 
the Schr\"{o}dinger equation for the quantum harmonic oscillator 
on the circle with energy eigenvalue given by 
$\lambda \equiv \pi^2 \epsilon/L^2+2 L^2/\pi^2$.

Suppose now we consider Mathieu's equation for
$K=L^4/\pi^4$ and $L$ asymptotic. In this case the
first levels are concentrated in values $y \ll L$ and
thus, according to previous discussion, these
energy levels of Mathieu's equation, which we call 
$\epsilon^L_n$, will provide the energy levels of 
the standard quantum
harmonic oscillator when $L \rightarrow \infty$.
Now, analogously to the definition of a quantum
particle through the ordinary quantum harmonic
oscillator, we define $n$ quantum particles 
on the circle of length $L$ as having energy $\epsilon^L_n$.
By consistency, $\epsilon^{L\rightarrow\infty}_n-
\epsilon^{L\rightarrow\infty}_0=n(\epsilon^{L\rightarrow\infty}_1-
\epsilon^{L\rightarrow\infty}_0)$.

In fact, there is a solution obtained by Ince and 
Goldstein 
$^{\cite{mathieu4}, \cite{mathieu5}, \cite{mathieu1}}$ to 
Mathieu's equation, Eq. (\ref{eq:mathieu1}), for 
asymptotic values of $K$. Their expansion for $\epsilon$,
the characteristic value of the equation, in the present case, 
i.e, $K=L^4/\pi^4$, provides for $\lambda$ the value
\begin{equation}
\lambda_n = \nu_n - \left( \nu_n^2 + 1 \right) \frac{a^2}{2^6} 
-\left( 5\nu_n^4+34\nu_n^2+9 \right) \frac{a^3}{2^{16}}+ \cdots \, ,
\label{eq:incegoldstein}
\end{equation}
where $\nu_n = 2 n +1$ and $a=\pi/L$. 
For $L \rightarrow \infty$,
($a\rightarrow 0$) we recognize the energy eigenvalues of the 
quantum harmonic oscillator. Thus we see that 
the above asymptotic solution $^{\cite{mathieu4}, \cite{mathieu5}, 
\cite{mathieu1}}$ of the characteristic values of
Mathieu's equation is a deformation of the quantum
harmonic oscillator with deformation parameter equal 
to $a=\pi/L$.

The above solution corresponds to the energy levels of
Mathieu's equation when the parameter $K$ appearing in
Eq. (\ref{eq:mathieu1}) is large, 
i.e, when $a^4 (2n+1)^2/16$ is small $^{\cite{mathieu4}, 
\cite{mathieu5}}$. Note that, even if $L$ is large 
which leads to a localization of the solution, this is 
periodic with period $2 L$.

Let us now consider dimensionfull variables. We call the
dimensionless variables $y$ and $L$ as $y \equiv x/x_0$ and
$L \equiv Z/x_0$, where $x$, $Z$ are dimensionfull and $x_0$ 
is a scale dimensionfull parameter. In this case when
the dimensionless variable $y$ varies from $-L$ to $L$ the
dimensionfull variable $x$ varies as $-Z \leq x \leq Z$. Thus,
$Z$ is the dimensionfull length of the one-dimensional
space. Furthermore, as explained before, the well known 
Schr\"{o}dinger's equation for the 
harmonic oscillator is obtained for $K=L^4/\pi^4$ in Eq. 
(\ref{eq:mathieu4}), when $L \rightarrow \infty$. In terms of
dimensionfull quantities this limit is achieved when
$Z >> x_0$. Therefore, we could say that $x_0$ is a scale where
deformed properties become relevant.

\section{Deformed Heisenberg algebra associated with Mathieu's
equation}     

The purpose of this section is to construct an algebra, like
the Heisenberg algebra, for the Mathieu system described in 
the previous section. Like the standard algebra for the 
quantum harmonic oscillator, the algebra we are going to
construct has creation and annihilation operators as part 
of its generators.

To this end let us consider an algebra generated by 
$J_{0}$, $A$ and $A^{\dagger}$ described by the relations 
$^{\cite{jpa}}$
\begin{eqnarray}
J_{0} \, A^{\dagger} &=& A^{\dagger} \, f(J_{0}) ,
\label{eq:alg1} \\
A \, J_{0} &=& f(J_{0}) \, A , 
\label{eq:alg2} \\
\left[ A, A^{\dagger} \right] &=& f(J_{0})-J_{0} ,
\label{eq:alg3}
\end{eqnarray}
where $^{\dagger}$ is the Hermitian conjugate and, 
by hypothesis, $J_{0}^{\dagger}=J_{0}$ and $f(J_{0})$
is a general analytic function of $J_{0}$. 

Using the algebraic relations in Eqs. (\ref{eq:alg1})-(\ref{eq:alg3}) we
see that the operator
\begin{equation}
C = A^{\dagger} \, A - J_{0} = A \, A^{\dagger} - f(J_{0})  
\label{eq:casimir}
\end{equation}
satisfies
\begin{equation}
\left[ C,J_{0} \right] = \left[ C,A \right] = 
\left[ C,A^{\dagger} \right] = 0  ,
\label{eq:comute}
\end{equation}
being thus a Casimir operator of the algebra.

We present now the representations of the algebra when 
the function $f(J_{0})$ is a general analytic function of  
$J_{0}$. 
We assume that we have an $n$-dimensional irreducible representation
of the algebra given by Eqs. (\ref{eq:alg1})-(\ref{eq:alg3}) and 
also that there is a state $|0\rangle$ with the lowest 
eigenvalue of the Hermitian operator $J_{0}$
\begin{equation}
J_{0} \, |0\rangle = \alpha_{0} \, |0\rangle .
\label{eq:alfa0}
\end{equation}
For each value of $\alpha_{0}$
we have a different vacuum and therefore a better notation 
for this state could be $|0\rangle_{\alpha_0}$. However, for 
simplicity, we shall omit subscript $\alpha_0$.  

Let $| m \rangle$ be a normalized eigenstate of $J_{0}$, 
\begin{equation}
    J_{0} |m \rangle = \alpha_{m} |m \rangle \, . 
    \label{eq:alfam}
\end{equation}
where  
\begin{equation}
\alpha_m = f^m(\alpha_0) = f(\alpha_{m-1}) \, ,
    \label{eq:alfam3}
\end{equation}
and $m$ denotes the number of iterations of $\alpha_{0}$ 
through $f$.  

As proved in \cite{jpa}, under the hypothesis stated previously
\footnote{$J_0$ is Hermitian and there exists a vacuum state.},
for a general function $f$ we obtain
\begin{eqnarray}
J_{0} \, |m\rangle &=& f^{m}(\alpha_0) \, |m\rangle , \; \; \; m = 0,1,2, 
\cdots \; , 
\label{eq:b1} \\
A^{\dagger} \, |m-1\rangle &=& N_{m-1} \, |m\rangle , 
\label{eq:b2} \\
A \, |m\rangle &=& N_{m-1} \, |m-1\rangle ,
\label{eq:b3}
\end{eqnarray}
where $N_{m-1}^2 = f^{m}(\alpha_0)-\alpha_0$. Note that
for each function $f(x)$ the representations are constructed
by the analysis of the above equations as done in
\cite{jpa} for the linear and quadratic $f(x)$.

When the functional $f(J_{0})$ is linear in $J_{0}$, i.e., 
$f(J_{0}) = q^2 J_0 +s$, it was shown in \cite{jpa} that the
algebra in Eqs. (\ref{eq:alg1})-(\ref{eq:alg3}) recovers the
$q$-oscillator algebra for $\alpha_0 = 0$.
Moreover, as shown in \cite{jpa}, where the representation
theory was constructed in detail for the linear and quadratic
functions $f(x)$, the essential tool to construct
representations of the algebra in (\ref{eq:alg1})-(\ref{eq:alg3})
for a general analytic function $f(x)$ is the analysis of the
stability of the fixed points of $f(x)$ and their composed
functions.

It was shown in \cite{jpa} and \cite{comhugo} that there is a class
of one-dimensional quantum systems described by these
generalized Heisenberg algebras. This class is characterized by
those quantum systems having energy eigenvalues given by
\begin{equation}
\varepsilon_{n+1} = f(\varepsilon_{n}) \, ,
\label{eq:class}
\end{equation}
where $\varepsilon_{n+1}$ and $\varepsilon_{n}$ are successive energy
levels and $f(x)$ is a different function for each physical
system. This function $f(x)$ is exactly the same function that
appears in the construction of the algebra in Eqs. 
(\ref{eq:alg1})-(\ref{eq:alg3}). In the algebraic description 
of this class of quantum systems, $J_0$ is the Hamiltonian 
operator of the system, $A^{\dagger}$ and $A$ are the creation 
and annihilation operators, respectively. This Hamiltonian and the 
ladder
operators are related by Eq. 
(\ref{eq:casimir}) where $C$ is the Casimir operator of the 
representation associated to the quantum system under
consideration.

Now, let us show that
the asymptotic solution to Mathieu's equation we
presented in the last section belongs to the class of 
algebras discussed previously. In other words, we shall 
construct a
Heisenberg-type algebra, an algebra with creation and 
annihilation operators, for the Ince-Goldstein solution (Eq. 
(\ref{eq:incegoldstein})) to the quantum
harmonic oscillator on $S^1$ and we
shall find the characteristic function $f(x)$ (see
Eqs. (\ref{eq:alg1})-(\ref{eq:alg3})) for this algebra. 
Furthermore,
we shall also propose a realization, as in the case
of the standard quantum harmonic oscillator, of the ladder 
operators in terms of the physical operators of the
system.

As described in \cite{comhugo} and \cite{campos}
the first thing we have to do in order to describe
the Heisenberg-type structure of a one-dimensional
quantum system is to relate the energy of the system
for two arbitrary successive levels (see Eq. (\ref{eq:class})). 
For the energy spectrum given in Eq. 
(\ref{eq:incegoldstein}), i.e,
\begin{equation}
\varepsilon_n^L = n + \frac{1}{2} - \left[ (2n+1)^2 + 1 \right] 
\frac{a^2}{2^7}
-\left[5(2n+1)^4+34(2n+1)^2+9 \right]\frac{a^3}{2^{17}}+ \cdots \, ,
\label{eq:incegoldstein2}
\end{equation}
we obtain 
\begin{equation}
\varepsilon_{n+1}^L = \varepsilon_{n}^L + 1 - \left( n+1 \right)\frac{a^2}{2^4} 
-(n+1)\left[ 10n(n+2)+21 \right]\frac{a^3}{2^{12}}+ \cdots \, .
\label{eq:incegoldstein3}
\end{equation}
Thus, we have to invert Eq. (\ref{eq:incegoldstein2}) in order
to obtain $n$ in terms of $\varepsilon_n^L$. Taking $n$ from Eq. 
(\ref{eq:incegoldstein2}) we get 
\begin{equation}
\varepsilon_{n+1}^L \equiv f(\varepsilon_n^L) = \varepsilon_{n}^L + 1 - 
(2\varepsilon_n^L+1)\frac{a^2}{2^5}- (2\varepsilon_n^L+1)
\left[ 20\varepsilon_n^L(\varepsilon_n^L+1)+27 \right]\frac{a^3}{2^{14}}
+ \cdots \, .
\label{eq:deff}
\end{equation}
According to Refs. \cite{mathieu4} and \cite{mathieu5}, this
solution is valid when $a^4 (2n+1)^2/16$ is small. Thus,
since $a=\pi/L$ is considered small, $n$ cannot be
very large.

Now, if we assume that $\varepsilon_n^L$ is the eigenvalue
of operator $J_0$ on state $|n\rangle$ we identify
$f(x)$ appearing in Eqs. (\ref{eq:b1})-(\ref{eq:b3})
with that one in Eq. (\ref{eq:deff}) for the quantum 
system under consideration. Then, the algebraic structure
describing the quantum system under consideration
is obtained using $f(x)$ defined in Eq. (\ref{eq:deff})
into Eqs. (\ref{eq:alg1})-(\ref{eq:alg3}) and can be written as
\begin{eqnarray}
\left[J_{0}, A^{\dagger}\right] &=& A^{\dagger} -
A^{\dagger} \, (2 J_{0}+1)\frac{a^2}{2^5}
-A^{\dagger}(2 J_{0}+1)\left[ 20 J_0(J_0+1)+27 \right] 
\frac{a^3}{2^{14}}+ \cdots , \nonumber \\
\label{eq:defheis1} \\
\left[ J_{0}, A \right] &=& -A +(2 J_{0}+1) \, A \frac{a^2}{2^5}
+(2 J_{0}+1)\left[ 20 J_0(J_0+1)+27 \right] \, A
\frac{a^3}{2^{14}}+ \cdots  ,\nonumber \\ 
\label{eq:defheis2} \\
\left[ A, A^{\dagger} \right] &=& 1-(2 J_{0}+1) \frac{a^2}{2^5}
-(2 J_{0}+1)\left[ 20 J_0(J_0+1)+27 \right]
\frac{a^3}{2^{14}}+ \cdots \,\, ,
\label{eq:defheis3}
\end{eqnarray}
where, according to Eqs. (\ref{eq:b1})-(\ref{eq:b3}), 
$A$ and $A^{\dagger}$ are the ladder operators for the 
system under consideration, i.e, $A^{\dagger}$ when
applied to state $|m\rangle$, that has $J_0$ 
eigenvalue $\varepsilon_m^L$, gives, apart from a 
multiplicative factor 
depending on $m$, the state $|m+1\rangle$ has
energy eigenvalue $\varepsilon_{m+1}^L$. A similar
role played by $A$.

Note that, when $a\rightarrow 0$ ($L\rightarrow \infty$),
we re obtain the well known Heisenberg algebra, as it
should be, since we showed in the previous section that
Mathieu's equation, Eq. (\ref{eq:mathieu1}), for
$K=L^4/\pi^4=a^{-4}$ gives the well known 
Schr\"{o}dinger's  equation for the harmonic
oscillator, Eq. (\ref{eq:schroed}), in this limit.

Now, let us realize the operators $A$, 
$A^{\dagger}$ and $J_0$ in terms of physical
operators as in the case of the one-dimensional
quantum harmonic oscillator, following what was done in
\cite{comhugo} and in \cite{campos} for the
square-well potential and $q$-oscillators
\cite{qft1}. 
Let us consider a  
one dimensional lattice in a momentum 
space where the 
momenta are allowed to take only discrete values, say $p_{0}$, 
$p_{0}+a$, $p_{0}+2a$, $p_{0}+3a$ etc, with $a>0$.
The left and right discrete  derivatives are given by 
\begin{eqnarray}
    (\partial_{p} \, f) \, (p) & = & \frac{1}{a} \, [f(p+a) - f(p)] \, ,
    \label{eq:partialleft}  \\
   (\bar{\partial}_{p} \, f) \, (p)  & = & \frac{1}{a} \, [f(p) - f(p-a)] \, , 
    \label{eq:partialright}
\end{eqnarray}
which are the two possible definitions of derivatives on a lattice.

Let us now introduce the momentum shift operators 
\begin{eqnarray}
    T  & = & 1 + a \, \partial_{p}
    \label{eq:a}  \\
    \bar{T} & = & 1 - a \, \bar{\partial}_{p} \, ,
    \label{eq:abarra}
\end{eqnarray}
which shift the momentum value by $a$
\begin{eqnarray}
    (Tf) \, (p) & = & f(p+a)
    \label{eq:af}  \\
    (\bar{T}f) \, (p) & = & f(p-a)
    \label{eq:abarraf}
\end{eqnarray}
and satisfies 
\begin{equation}
    T \, \bar{T} = \bar{T} T = \hat{1} \, ,
    \label{eq:aabarra}
\end{equation}
where $\hat{1}$ means the identity on the algebra of functions of $p$.  

Introducing the momentum operator $P$$^{\cite{dimakis1}}$
\begin{equation}
    (Pf) \, (p) = p \, f(p) \, ,
    \label{eq:momentum}
\end{equation}
we have 
\begin{eqnarray}
    T P & = & (P+a)T
    \label{eq:ap}  \\
    \bar{T} P & = & (P-a) \bar{T} \, \, .
    \label{eq:abarrap}
\end{eqnarray} 

Now, we go back to the realization of the deformed
Heisenberg algebra Eqs. 
(\ref{eq:defheis1})-(\ref{eq:defheis3}) in terms of 
physical operators. We can associate to the crystalline
structure of Mathieu's equation discussed in the
previous section the one dimensional lattice we have
just presented.

Observe that we can write $J_0$ for the asymptotic 
Ince-Goldstein solution to Mathieu's equation, 
Eq. (\ref{eq:incegoldstein2}), as
\begin{equation}
J_0 = \frac{P}{a} + \frac{1}{2} -  
\left[ \left( 2\frac{P}{a}+1 \right)^2 + 1 \right]
\frac{a^2}{2^7} - 
\left[5(2\frac{P}{a}+1)^4+34(2\frac{P}{a}+1)^2+
9 \right]\frac{a^3}{2^{17}}
+ \cdots  \, \, ,
\label{eq:defj0}
\end{equation} 
where $P$ is given in Eq. (\ref{eq:momentum}) and
its application to the vector states $|m\rangle$
appearing in (\ref{eq:b1})-(\ref{eq:b3}) gives
\begin{equation}
P \, |m\rangle = m \, a \, |m\rangle \,\, , m=0,1, 
\cdots  \, \, ,
\label{eq:aplicmom}
\end{equation}
and
\begin{equation}
\bar{T} \, |m\rangle = |m+1\rangle \,\, , m=0,1, 
\cdots  \, \, ,
\label{eq:aplictbar}
\end{equation}
where $\bar{T}$ and $T=\bar{T}^{\dagger}$ are defined
in Eqs. (\ref{eq:a})-(\ref{eq:aabarra}). It is useful to note
that from Eq. (\ref{eq:aplicmom}) it is possible to 
define the number operator $N$ as $N \equiv P/a$.

With the definition of $J_0$ given in Eq. (\ref{eq:defj0})
we see that $\varepsilon_n^L$ given in Eq. 
(\ref{eq:incegoldstein3}) is the $J_0$ eigenvalue of
state $|n\rangle$ as desired. Let us now define
\begin{eqnarray}
A^{\dagger}  &=& S(P) \, \bar{T} \,\,  ,
\label{eq:real1} \\
A &=& T \, S(P) \,\,  ,
\label{eq:real2} 
\end{eqnarray}
where,
\begin{equation}
S(P)^2 = \frac{P}{a} - \left[ \left( 2\frac{P}{a}+1 
\right)^2 -1 \right] \frac{a^2}{2^7}
-\left[5\left( 2\frac{P}{a}+1 \right)^4+34\left( 2
\frac{P}{a}+1\right)^2-
39 \right]\frac{a^3}{2^{17}}+ 
\cdots  ,
\label{eq:defS}
\end{equation}
satisfies $S^2(P) = J_0 - \alpha_0$ where $\alpha_0$,
defined in Eq. (\ref{eq:alfa0}), is $\varepsilon_0^L$.
Following Ref. \cite{circulo} one can show that
$A^{\dagger}$, $A$ and $J_0$ given in Eqs. (\ref{eq:real1}),
(\ref{eq:real2}) and (\ref{eq:defj0}) respectively, obey
the algebra defined in Eqs. (\ref{eq:defheis1}),
(\ref{eq:defheis2}) and (\ref{eq:defheis3}).

Note that the realization we have found in 
Eqs. (\ref{eq:real1}), (\ref{eq:real2}) and 
(\ref{eq:defj0}) is qualitatively different from the 
the realization of the standard
harmonic oscillator. This is reasonable,
since we have two physically different systems. Even
if the standard quantum harmonic oscillator defined on
$-\infty \leq x \leq \infty$ is a limiting case of the
periodic one, it is not periodic and in this case there
is no lattice
associated to it. On the other hand, once $L$ is finite,
$-L \leq x \leq L$, the periodic structure is explicitly
manifest and the realization in the finite case, given
in Eqs. (\ref{eq:real1}), (\ref{eq:real2}) and 
(\ref{eq:defj0}), shows it clearly.

\section{A quantum field theory in a box}

In section II we have presented a description
of a quantum harmonic oscillator on a circle and
in section III, its associated Heisenberg-type 
algebra, i.e., an algebra having the Hamiltonian 
and the step operators as generators, corresponding 
to a quantum harmonic oscillator on a circle. This algebra 
is a deformed Heisenberg algebra which goes to the 
standard Heisenberg algebra when the radius of the 
circle goes to infinite.

In this section, using the hypothesis that the successive
energy levels of the quantum harmonic oscillator on a circle
are still obtained by the creation or/and annihilation
of a quantum particle on a periodic structure,  we are 
going to construct a quantum field theory in a compact
space.  

In the momentum space appropriated to the realization of
the deformed Heisenberg algebra we discussed, besides
the operator $P$ defined in Eq. (\ref{eq:momentum}),
one can define two self adjoint operators as
\begin{eqnarray}
    \chi & \equiv & - i \left( S(P)(1-a \bar{\partial}_p )- 
(1+a \partial_p )S(P) \right) =
- i(A - A^{\dagger})  \, \, ,
    \label{eq:cord1}  \\
     Q & \equiv &  S(P)(1-a \bar{\partial}_p )+ 
(1+a \partial_p )S(P) =
A + A^{\dagger}  \, \, ,
    \label{eq:cord2}  
\end{eqnarray}
where $\partial_p$ and $\bar{\partial}_p$ are the left and right 
discrete derivatives defined in Eqs. (\ref{eq:partialleft}), 
(\ref{eq:partialright}).

It can be verified that operators $P$, $\chi$ and $Q$ generate
the following algebra on the momentum lattice:
\begin{eqnarray}
\left[ \chi,P \right] &=&  i a Q  ,
\label{eq:fecho1} \\
\left[ P,Q \right] &=&  i a \chi , 
\label{eq:fecho2} \\
\left[ \chi,Q \right] &=& 2 i S(P) \left( S(P+a)-S(P-a) \right) .
\label{eq:fecho3}
\end{eqnarray}
This algebra is the analog of the Heisenberg algebra
in the deformed case.

Since the analog of the Heisenberg algebra for the
deformed case has three generators, it is convenient
to define three
fields which we call $\phi(\vec{r},t)$, $\Pi(\vec{r},t)$
and $\wp(\vec{r},t)$.  In terms of Fourier series
these fields are given as
\begin{eqnarray}
\phi(\vec{r},t) &=&  \sum_{\vec{k}} 
\frac{1}{\sqrt{2\Omega\omega(\vec{k})}}  Q_{\vec{k}}(t) 
\, e^{i \vec{k}.\vec{r}} ,
\label{eq:defcampo1} \\
\Pi(\vec{r},t) &=&  \sum_{\vec{k}} 
\frac{i\omega(\vec{k})}{\sqrt{2\Omega\omega(\vec{k})}} 
\chi_{-\vec{k}}(t) 
\, e^{i \vec{k}.\vec{r}} , 
\label{eq:defcampo2} 
\end{eqnarray}
where $\omega(\vec{k})= \sqrt{\vec{k}^2+m^2}$, $m$ 
is a real parameter
and $\Omega$ is the volume of a rectangular box and
\begin{equation}
\wp(\vec{r},t) = \sum_{\vec{k}} \sqrt{\frac{\omega(\vec{k})}
{2\Omega}}  
\, \, \, S_{\vec{k}}  \, e^{i \vec{k}.\vec{r}}  \, .
\label{eq:defcampo3}  
\end{equation}
The time-dependent operators in the Hilbert space $Q_{\vec{k}}(t)$, 
$\chi_{\vec{k}}(t)$ and $S_{\vec{k}}$ 
will be defined in what follows and the components of 
$\vec{k}$ are given by
\begin{equation}
k_i = \frac{2 \pi l_i}{Z_i} ,\, \,\,\,  i=1,2,3 \,\,\,\,\,\,\, ,
\label{eq:kspace}
\end{equation}
with $l_i= 0,\pm 1,\pm 2, \cdots $ and $Z_i$ being the lengths of the 
three sides of a rectangular box $\Omega$. We introduce for
each point of this $\vec{k}$-space an independent deformed quantum
harmonic oscillator constructed in the last two previous
sections such that the deformed operators commute for different
three-dimensional lattice points. 
We also introduce an independent copy of the one-dimensional
momentum lattice defined in the previous section  
for each point of this $\vec{k}$-lattice  
so that $P_{\vec{k}}^{\dagger} = P_{\vec{k}}$ and 
$T_{\vec{k}}$, $\bar{T}_{\vec{k}}$ and $S_{\vec{k}}$ 
are defined by means of the previous definitions,
Eqs. (\ref{eq:a})-(\ref{eq:abarra}) and (\ref{eq:defS}),
through the substitution $P \rightarrow P_{\vec{k}}$.

It is possible to show that
\begin{eqnarray}
    A^{\dagger}_{\vec{k}} & = &  S_{\vec{k}} \, \bar{T}_{\vec{k}} \, ,
    \label{eq:j+k}  \\
     A_{\vec{k}} & = & T_{\vec{k}} \, S_{\vec{k}} \, ,    
    \label{eq:j-k}  \\
    J_0(\vec{k}) &=& \frac{P_{\vec{k}}}{a} + \frac{1}{2} -  
\left[ \left( 2\frac{P_{\vec{k}}}{a}+1 \right)^2 + 1 \right]
\frac{a^2}{2^7} - \nonumber \\
& &
\left[5(2\frac{P_{\vec{k}}}{a}+1)^4+34(2\frac{P_{\vec{k}}}{a}+1)^2+
9 \right]\frac{a^3}{2^{17}}
+ \cdots  \, \, ,
    \label{eq:j0k}
\end{eqnarray}
where
\begin{equation}
S_{\vec{k}}^2 = \frac{P_{\vec{k}}}{a} - \left[ 
\left( 2\frac{P_{\vec{k}}}{a}+1 
\right)^2 -1 \right] \frac{a^2}{2^7}
-\left[5\left( 2\frac{P_{\vec{k}}}{a}+1 \right)^4+34\left( 2
\frac{P_{\vec{k}}}{a}+1\right)^2-
39 \right]\frac{a^3}{2^{17}}+ 
\cdots  ,
\label{eq:defSk}
\end{equation}
satisfy the algebra in Eqs. (\ref{eq:defheis1}),
(\ref{eq:defheis2}) and (\ref{eq:defheis3})
for each point of this $\vec{k}$-lattice and the operators
$A^{\dagger}_{\vec{k}}$, $A_{\vec{k}}$ and $J_{0}(\vec{k})$ 
commute among them for different points of this
$\vec{k}$-lattice.

Now, we define operators $\chi$ and $Q$ for
each point of the three-dimensional lattice as
\begin{eqnarray}
    \chi_{\vec{k}} & \equiv & -i(T_{-\vec{k}} \, S_{-\vec{k}} - 
S_{\vec{k}} \, \bar{T}_{\vec{k}} ) =
- i ( A_{-\vec{k}} - A^{\dagger}_{\vec{k}})   \, \, ,
    \label{eq:cord5}  \\
     Q_{\vec{k}} & \equiv &  T_{\vec{k}} \, S_{\vec{k}} + 
S_{-\vec{k}} \, \bar{T}_{-\vec{k}} =
A_{\vec{k}} + A^{\dagger}_{-\vec{k}}  \, \, ,
    \label{eq:cord6}  
\end{eqnarray}
such that $\chi_{\vec{k}}^{\dagger}= \chi_{-\vec{k}}$ and 
$Q_{\vec{k}}^{\dagger}= Q_{-\vec{k}}$, exactly as it happens
in the construction of a spin-$0$ field for the spin-$0$ quantum 
field theory $^{\cite{tdlee}}$. These operators appear in the 
Fourier expansion of the fields given in Eqs. 
(\ref{eq:defcampo1})-(\ref{eq:defcampo3}).

By a straightforward calculation, one can show that the 
Hamiltonian
\begin{eqnarray}
H = \frac{1}{2} \int_{\Omega} d^3 r \left(  \Pi(\vec{r},t)^2
 +\rho \, | \wp(\vec{r},t)|^2 +
   \phi(\vec{r},t)  (-{\vec{\nabla}}^2+m^2)  
\phi(\vec{r},t)   \right)  \,\, ,
\label{eq:defhamilt}
\end{eqnarray}
can be written as
\begin{eqnarray}
H &=& \frac{1}{2} \sum_{\vec{k}} \omega(\vec{k}) 
\left( A^{\dagger}_{\vec{k}} A_{\vec{k}}+A_{\vec{k}} 
A^{\dagger}_{\vec{k}}+\rho \, S(N_{\vec{k}})^2 \right) \nonumber \\
&=&\frac{1}{2}\sum_{\vec{k}} \omega(\vec{k}) \left( S(N_{\vec{k}}+1)^2 +
(1+\rho) \, S(N_{\vec{k}})^2 \right)   \,\,  ,
\label{eq:resulthamilt}
\end{eqnarray}
where $\rho$ is an arbitrary number and 
\begin{equation}
S(N_{\vec{k}})^2 = N_{\vec{k}} - \left[ 
\left( 2N_{\vec{k}}+1 
\right)^2 -1 \right] \frac{a^2}{2^7}
-\left[5\left( 2N_{\vec{k}}+1 \right)^4+34\left( 2
N_{\vec{k}}+1\right)^2-
39 \right]\frac{a^3}{2^{17}}+ 
\cdots  , \, \, .
\label{eq:defs2}
\end{equation}
Since the term in the Hamiltonian, (\ref{eq:defhamilt}), proportional
to $\rho$ is time-independent, it seems that it cannot produce any 
relevant effect. Thus, for simplicity, we will take $\rho=0$.
In order that the energy of the vacuum state becomes zero
we replace $H$ in Eq. (\ref{eq:resulthamilt}) by
\begin{equation}
H = \frac{1}{2}\sum_{\vec{k}} \omega(\vec{k}) \left( S(N_{\vec{k}}+1)^2 +
S(N_{\vec{k}})^2 - N_0^2 \right) \,\, 
\label{eq:resulthamilt1}
\end{equation}
where 
\begin{equation}
N_0^2 \equiv f(\alpha_0)-\alpha_0=1-a^2/2^4-21 a^3/2^{12}
+\cdots \;.
\label{eq:enezero} 
\end{equation}
Note that in the limit $L \rightarrow \infty$,
the above Hamiltonian is proportional to the number operator.

The eigenvectors of $H$ form a complete set and span
the Hilbert space of this system. They are the following
\begin{equation} 
|0 \rangle, \,\, A^{\dagger}_{\vec{k}} |0 \rangle, \,\,
A^{\dagger}_{\vec{k}} A^{\dagger}_{\vec{k}'} |0 \rangle \,\,
\mbox{for} \,\, \vec{k}\not= \vec{k}', \,\,
(A^{\dagger}_{\vec{k}})^2 |0 \rangle, \,\, \cdots \,\, ,
\label{eq:hilbert} 
\end{equation}
where the state $|0\rangle$ satisfies as usual 
$A_{\vec{k}} |0\rangle =0$ (see Eq. 
(\ref{eq:alg3})) for all $\vec{k}$ and $A_{\vec{k}}$, 
$A^{\dagger}_{\vec{k}}$ for each $\vec{k}$
satisfying the deformed Heisenberg algebra 
Eqs. (\ref{eq:defheis1})-(\ref{eq:defheis3}).

The time evolution of the fields can be studied by
means of Heisenberg's equation for $A^{\dagger}_{\vec{k}}$,
$A_{\vec{k}}$ and $S_{\vec{k}}(\equiv S(N_{\vec{k}}))$. 
Define
\begin{equation}
h(N_{\vec{k}})  \equiv  \frac{1}{2} \left(
 S^2(N_{\vec{k}}+2) - 
 S^2(N_{\vec{k}}) \right) \,\, .
\label{eq:defh}  
\end{equation}
Thus, using Eq. (\ref{eq:resulthamilt}) or (\ref{eq:resulthamilt1})
and $\left[ N,A^{\dagger}\right]=A^{\dagger}$ we obtain
\begin{equation}
\left[ H, A^{\dagger}_{\vec{k}} \right] =
\omega(\vec{k}) \, A^{\dagger}_{\vec{k}} \,\, h(N_{\vec{k}})   \,\, .
\label{eq:comutheis1}  
\end{equation}
We can solve Heisenberg's equation for the deformed case
and the result is
\begin{equation}
A^{\dagger}_{\vec{k}}(t) =  A^{\dagger}_{\vec{k}}(0) \,\,
e^{i \omega(\vec{k}) \, h(N_{\vec{k}}) \, t}   \,\, .
\label{eq:solveheis1}  
\end{equation}
Note that for $L \rightarrow \infty$ we have 
$h(N_{\vec{k}}) \rightarrow 1$ and Eq. (\ref{eq:solveheis1})
gives the correct result for this undeformed case.
Furthermore, we easily see 
that operators $P_{\vec{k}}$ and $S_{\vec{k}}$ are 
time-independent. We emphasize that the extra term, $h(N_{\vec{k}})$,
in the exponentials depends on the number operator, being this the
main difference from the undeformed case. The Fourier
transformation of Eq. (\ref{eq:defcampo1}) can then be written as
\begin{equation}
\phi(\vec{r},t) = \alpha(\vec{r},t) + \alpha(\vec{r},t)^{\dagger}
 \,\, ,
\label{eq:defcampotempo}  
\end{equation}
where
\begin{equation}
\alpha(\vec{r},t) =  \sum_{\vec{k}} 
\frac{1}{\sqrt{2\Omega\omega(\vec{k})}} \,\,   
e^{i \vec{k}.\vec{r}-i \omega(\vec{k}) \,
h(N_{\vec{k}}) \, t} A_{\vec{k}}  \, \,\, ,
\label{eq:defcampotempo1}  
\end{equation}
with $A_{\vec{k}}$ in Eq. (\ref{eq:defcampotempo1}) being  
time-independent and $\alpha(\vec{r},t)^{\dagger}$
is the Hermitian conjugate of $\alpha(\vec{r},t)$.

The Dyson-Wick contraction between 
\footnote{$x_i \equiv (\vec{r_i},t_i)$}
$\phi(x_1)$ and $\phi(x_2)$,
can be computed using Eqs.
(\ref{eq:defcampotempo})-(\ref{eq:defcampotempo1}),
which results in 
\begin{equation}
D^{N}_F(x_1,x_2) = \sum_{\vec{k}} \frac{e^{i\vec{k} . \Delta 
\vec{r}_{12}}}{2\Omega\omega(\vec{k})}
\left( S(N_{\vec{k}}+1)^2 \, e^{\mp i \omega(\vec{k})\, h(N_{\vec{k}})\,
\Delta t_{12}} - S(N_{\vec{k}})^2 \, e^{\mp i \omega(\vec{k})\, 
h(N_{\vec{k}}-1)\, \Delta t_{12}} 
\right) \, ,
\label{eq:propsum}
\end{equation}
where $\Delta t_{12}=t_1-t_2$, $\Delta \vec{r}_{12}=
\vec{r}_1-\vec{r}_2$. The minus sign in the exponent
holds when $t_1 > t_2$ and the positive sign when
$t_2 > t_1$. Note that when $L \rightarrow \infty$, 
$h(N_{\vec{k}}) \rightarrow 1$ and $S(N_{\vec{k}}+1)^2 -
S(N_{\vec{k}})^2 \rightarrow 1$ recovering the 
standard result for the propagator.

We shall now present the result concerning the perturbative
computation of the first order scattering process 
$1+2 \rightarrow 1^{'}+2^{'}$ for $p_1 \not= p_2 \not=
p_1' \not= p_2'$ with the initial state
\begin{equation}
|1,2 \rangle \equiv \frac{1}{N_0^2} A^{\dagger}_{p_1} 
\, A^{\dagger}_{p_2} |0\rangle
\,\, , 
\label{eq:estinic}
\end{equation} 
and the final state
\begin{equation}
|1^{'},2^{'} \rangle \equiv \frac{1}{N_0^2} A^{\dagger}_{p_1^{'}} \, 
A^{\dagger}_{p_2^{'}} |0\rangle
\,\, , 
\label{eq:estfin}
\end{equation}
where $A_{p_i}$ and $A^{\dagger}_{p_i}$ satisfy the algebraic 
relations in Eqs. (\ref{eq:defheis1})-(\ref{eq:defheis3}).These
particles are supposed to be described by the Hamiltonian
given in Eq. (\ref{eq:defhamilt}) with an interaction
given by $\lambda \int_{\Omega_t} :\phi(\vec{r},t)^4: d^3r$,
where $\Omega_t=\Omega \otimes t$ is the four volume of
integration. To the
lowest order in $\lambda$, we have ($\Gamma$ means the
standard $S$-matrix)
\begin{equation}
\langle 1^{'},2^{'} | \Gamma | 1,2 \rangle_1 = -i \lambda 
\int_{\Omega_t} 
d^4 x \langle 1^{'},2^{'} | :\phi^4(x): | 1,2 \rangle \,\, .
\label{eq:matrixelem1}
\end{equation}

The first order computation follows, step by step, the
computation of the first order scattering process given in Ref.
\cite{qft1} and gives us the following result
\begin{eqnarray}
&&\langle 1^{'},2^{'} | \Gamma | 1,2 \rangle_1 = \frac{-6i(2\pi)^4 
}{\Omega^2 \sqrt{\omega_{\vec{p}_1}\omega_{\vec{p}_2} 
\omega_{\vec{p'}_1}\omega_{\vec{p'}_2} }}\frac{\lambda{N_0}^4}
{h(0)}  
\delta^4(P_{1}+P_{2}-P'_{1}-P'_{2})  \,\, ,  \nonumber \\
\label{eq:finalmatrix1}
\end{eqnarray}
where
\begin{eqnarray}
P_{i} = \left(\vec{p}_i, \omega_{\vec{p}_i}
\right), \; 
P'_{i} = \left(\vec{p'}_i, \omega_{\vec{p'}_i}
\right)   
\nonumber \\
\label{eq:definePs}
\end{eqnarray}
and from Eq. (\ref{eq:defh})
\begin{equation}        
         h(0) =  1-3 \frac{a^2}{2^5}-
        123 \frac{a^3}{2^{13}} + \ldots  \;\; .
\label{eq:agazero}
\end{equation} 
Note that 
when $L \rightarrow \infty \; (a \rightarrow 0)$ we have 
$N_0 \rightarrow 1$, $h(0) \rightarrow 1$, the box 
$\Omega$ becomes an infinite box and Eq. (\ref{eq:finalmatrix1}) 
becomes the standard undeformed result $^{\cite{tdlee}}$. It 
is convenient to note at this point that we are, by hypothesis, 
identifying the linear dimensions of the box $\Omega$ where we 
perform the spatial integration in Eq. (\ref{eq:matrixelem1}) 
with the dimensionfull length, $Z$ ($L=Z/x_0$), of the circle 
where the 
harmonic oscillator is defined. This identification is 
not strictly necessary, it comes from our approach
that everything happens inside the spatial box $\Omega$. 
We suppose that in a universe approximated by a finite 
spatial box $\Omega$ 
the step 
operators of the quantum harmonic oscillator defined on a
circle of length $Z$, where $Z$ is a linear dimension  
of the box $\Omega$, create and/or annihilate point 
particles. Thus, if the spatial box $\Omega$ increases so 
does the length of the circle where the harmonic   
oscillator is defined.

To second order in $\lambda$ the scattering process 
$1+2 \rightarrow 1^{'}+2^{'}$ is given as
\begin{equation}
\langle 1^{'},2^{'} | \Gamma | 1,2 \rangle_2 = \frac{(-i)^2}{2} 
\lambda^2 \int \int_{\Omega_t} d^4 x \, d^4 y 
\langle 1^{'},2^{'} | 
T(:\phi^4(x): :\phi^4(y):) | 1,2 \rangle \,\, ,
\label{eq:matrixelem2}
\end{equation}
where $T$ denotes the time-ordered product. In order to
convert the time-ordered product into a normal product
we use the Wick's expansion.
The propagator in the present case, see Eq. 
(\ref{eq:propsum}), is not a simple $c$-number since 
it depends on the number operator $N$ and this fact 
induces modifications in the standard Wick 
expansion. This subject was already discussed in \cite{qft2}
where the computation of a scattering process for a 
deformed QFT to second order in the coupling constant
was presented. 

Following Ref. \cite{qft2} we find for the scattering
process under consideration, up to second order in the
coupling constant, is given by
\begin{eqnarray}
\langle 1^{'},2^{'} | \Gamma | 1,2 \rangle_2 = \frac{1}
{2\Omega^2\sqrt{\omega_{\vec{p}_1}\omega_{\vec{p}_2} 
\omega_{\vec{p'}_1}\omega_{\vec{p'}_2}}} 
\left(\frac{\lambda{N_0}^4}{h(0)}\right)^2
\delta^4(P_{1}+P_{2}-P'_{1}-P'_{2}) \nonumber \\
\left( I+I'+I''+I''' \right) \, ,
\label{eq:resmatrixelem2}
\end{eqnarray}
where 
\begin{equation}
I = -\frac{(2\pi)^2}{4\Omega} \sum_{\vec{k}}
\frac{1}{\sqrt{(\vec{k}^2+m^2)\left[(\vec{s}-\vec{k})^2
+m^2\right]}} \, ,
\label{eq:defI}
\end{equation}
with $\vec{s}=\vec{p}_1+\vec{p}_2$ and
\begin{eqnarray}
 I' &=& I(\vec{s}\rightarrow -\vec{s}) \, , \\
 I'' &=& I(\vec{s}\rightarrow \vec{t}\equiv 
 \vec{p}_1-\vec{p'}_1) \, , \\ 
 I''' &=& I(\vec{s}\rightarrow \vec{u}\equiv 
 \vec{p}_1-\vec{p'}_2) \, .
\label{eq:defdifI}
\end{eqnarray}

In summary, the scattering process 
$1+2 \rightarrow 1^{'}+2^{'}$ for $p_1 \not= p_2 \not=
p_1' \not= p_2'$ with the initial
and final states given in Eqs. (\ref{eq:estinic}), 
(\ref{eq:estfin}) respectively,
where $A_{p_i}$, $A^{\dagger}_{p_i}$ satisfy the 
algebraic 
relations in Eqs. (\ref{eq:defheis1})-(\ref{eq:defheis3})
and the particles are supposed to be described by 
the Hamiltonian
given in Eq. (\ref{eq:defhamilt}) with an interaction
given by $\lambda \int_{\Omega_t} 
:\phi(\vec{r},t)^4: d^3r$ is 
given up to second order in the coupling constant,
$\lambda$, as
\begin{equation}
\langle 1^{'},2^{'} | \Gamma | 1,2 \rangle = 
\frac{\lambda N_0^4}{h(0)} A_1 + \left( \frac{\lambda N_0^4}
{h(0)}\right)^2 \left( A_2^s + A_2^t + A_2^u \right) \; ,
\label{eq:finres}
\end{equation}
where $A_1$ is obtained from Eq. (\ref{eq:finalmatrix1}), 
$A_2^s$ comes from $I$ and $I'$ in Eq. (\ref{eq:resmatrixelem2}), 
$A_2^t$ and $A_2^u$ come from $I''$ and $I'''$ respectively.

Note that 
when $L \rightarrow \infty \; (a \rightarrow 0)$ we have 
$N_0 \rightarrow 1$, $h(0) \rightarrow 1$, the box 
$\Omega$ becomes an infinite box, Eq. (\ref{eq:finres}) 
becomes the standard undeformed result with $A_1$, $A_2^s$, 
$A_2^t$ and $A_2^u$ being the same contributions that we 
find in the {\it standard} $\lambda$ $\phi^4$ (non-deformed) 
model corresponding to the tree level, the $s$, $t$ and $u$ 
channels for one-loop level, respectively. Also, it is worth
to notice that
the perturbative expansion shows that the coupling constant 
which appear in the interacting Hamiltonian is modified
as $\lambda \rightarrow \lambda N_0^4/h(0)$. This means that
the effective coupling constant, $\lambda_{eff}\equiv \lambda 
N_0^4/h(0)$, in this 
framework is modified due to the presence of the deformation 
parameter $a=\pi/L$.

\section{Contribution from the boundary for the 
variation of the coupling constant}

The comments in the last paragraph of the previous section allow us
to connect the effective coupling constant appearing in the
perturbation expansion, which is given by $\lambda N_0^4/h(0)$,
with the size of the box we are considering, i.e.,
the linear dimension $Z$ of the box $\Omega$. In this section, 
based on this connection we are going to compute the variation 
of the effective coupling constant for two different values of $Z$, 
namely one corresponding to the time of nucleosynthesis of the
standard cosmological model and the other to the present epoch. 
The choice of these values, as said before, was done just to perform a 
calculation. With this choice we are not assuming that the Universe is 
described by our model. In fact, we want just to have an idea of what 
would be the contribution from the boundary, in a compact space, to 
the variation of the coupling constant in the framework of our approach.

In order to investigate the variation of the effective coupling
constant,
let us define $p \equiv N_0^4/h(0)$ which for two different values
of $L$, namely $L_{\pm}$ gives
\begin{equation}
 p_{\pm} = \frac{\left(1-a_{\pm}^2/2^4 - 21 a_{\pm}^3/2^{12} +\cdots 
 \right)^2}{1-3 a_{\pm}^2/2^5-
        123 a_{\pm}^3/2^{13} + \ldots}        \; ,
\label{eq:pmaismenos}
\end{equation}
where we have used Eqs. (\ref{eq:enezero}) and
(\ref{eq:agazero}) with
$a_{\pm}=\pi/L_{\pm}$, $L_{\pm}=Z_{\pm}/x_0$ and
$\lambda_{eff}^{\pm}= \lambda p_{\pm}$.

In what follows let us compute the dimensionless quantity,
$\Delta\alpha/\alpha$, given by
\begin{equation}
\frac{\Delta\alpha}{\alpha}= \frac{(\lambda_{eff}^{+})^2-
(\lambda_{eff}^{-})^2}{(\lambda_{eff}^{+})^2} =
1-\frac{(\lambda_{eff}^{-})^2}{(\lambda_{eff}^{+})^2} \; ,
\label{eq:delta1}
\end{equation}
where $\pm$ means the present time and the time at the moment
of nucleosynthesis, respectively.

We have assumed that the creation
and/or annihilation operators of the quantum harmonic oscillator on a
periodic line creates and/or annihilates a quantum particle.
Along these lines we showed in Sections II and III that there
is a deformation parameter $a$ which is connected to the linear
size of the box where the second quantized formalism is constructed
through $a=\pi/L$ with $L$ given by $L=Z/x_0$. As discussed in Section II, 
$x_0$ is a scale where the deformation starts to become relevant.
In what follows, we will assume that the value of $x_0$ is at
least equal to the one corresponding to the scale of the electroweak
phase transition just to have a reference size which will permit to perform 
our calculation . 

Now, let us compute the dimensionless quantity
$\Delta\alpha/\alpha$ for two different values of the dimensions
of the box, namely, for $Z_+ \approx 10^{28} cm$ and
$Z_- \approx 10^{19} cm$ in accordance with our choice .
For these values of the size of the box, the 
deformation parameters are $a_+=\pi/L_+=\pi x_0/Z_+ \approx
10^{-15}$ and $a_- \approx 10^{-6}$. Due to the magnitude of
the deformation parameters under consideration it is suffice to 
take the lowest order expansion for $p_{\pm}$, which is
\begin{equation}
p_{\pm} = 1- 
\frac{a_{\pm}^2}{2^5}+\cdots \; .
\label{eq:pmaismenosexpand}
\end{equation}
Taking into account this expansion and the estimated value
of $a_-$ we obtain for $\Delta\alpha/\alpha$ the following result 
\begin{equation}
\frac{\Delta\alpha}{\alpha} =  
\frac{a_{-}^2}{2^5}+\cdots  <  10^{-12} \; .
\label{eq:estimativa}
\end{equation}

\section{Final comments}
In this paper we have constructed a QFT in a finite box.
In order to construct this QFT, we used the hypothesis that
particles living in a finite box with periodic boundary
conditions are created and/or annihilated through the 
creation and/or annihilation operators, respectively, 
of a quantum harmonic oscillator on a circle.

The quantum harmonic oscillator we have used is described
by Mathieu's equation and its associated creation and 
annihilation operators obey a deformed Heisenberg algebra. We have
thus followed the treatment given in Refs. \cite{qft1} 
and \cite{qft2}, which presents the construction of a
deformed QFT based on $q$-oscillators,
in order to construct the present QFT in a finite 
box.

The perturbative series we have found shows an effective
coupling constant given by $\lambda_{eff}= \lambda 
N_0^4/h(0)$, where $N_0$ and $h(0)$, see Eqs. 
(\ref{eq:enezero}) and (\ref{eq:agazero}), 
which depends on the dimensionless quantity 
$L=Z/x_0$, where $Z$ is  a linear dimension of the finite
box $\Omega$ and $x_0$ is a scale where the modified
description of the creation of particles starts to be
relevant.

Even though our model is not a cosmological model, we considered
an estimation of the bound for the variation of the effective
coupling constant taking into account two
values, $Z$, of the linear dimension of the finite box $\Omega$, 
one corresponding to the time of the nucleosynthesis of the 
standard cosmological model and the other to the size of the
Universe nowadays. The result obtained, $\Delta\alpha/\alpha 
< 10^{-12}$, is in the range of results on the 
constraints for the variation of the coupling constants for
different epochs of Universe
\cite{varyingc}, we thus think that it would be interesting
to analyze the approach we considered in this paper in a
cosmological model.

\vspace{0.1 cm}

\noindent
{\bf Acknowledgments:}
The authors thank CNPq and PRONEX/CNPq for partial support.

\newpage


\begin{thebibliography}{30}
\bibitem{tdlee} See for instance: T. D. Lee, ``Particle Physics and
Introduction to Field Theory", Harwood academic publishers,
New York, 1981. 
\bibitem{circulo} M. A. Rego-Monteiro, Eur. Phys. J. {\bf C 21}
(2001) 749.
\bibitem{qft1} V. B. Bezerra, E. M. F. Curado and M. A. Rego-Monteiro, 
Phys. Rev. {\bf D 65} (2002) 065020.
\bibitem{qosc} A. J. Macfarlane, J. Phys. {\bf A 22} (1989) 4581;
L. C. Biedenharn, J. Phys. {\bf A 22} (1989) L873.
\bibitem{qft2} V. B. Bezerra, E. M. F. Curado and M. A. Rego-Monteiro, 
Phys. Rev. {\bf D 66} (2002) 085013.
\bibitem{qft3} V. B. Bezerra, E. M. F. Curado and M. A. Rego-Monteiro, 
Phys. Rev. {\bf D 69} (2004) 085003.
\bibitem{ok1} Y. Ohnuki and S. Kitakado, J. Math. Phys.
{\bf 34} (1993) 2827.
\bibitem{ok2} S. Tanimura, Prog. Theor. Phys. {\bf 90} (1993) 271;
K. Takenaga, Phys. Rev. {\bf D62} (2000) 065001.
\bibitem{mathieu1} R. Campbell, ``Th\'{e}orie G\'{e}n\'{e}rale
de L' \'{E}quation de Mathieu", Masson et Cie. \'{e}diteurs, Paris, 1955.
\bibitem{mathieu2} E. H. Lieb and D. C. Mattis, ``Mathematical
Physics in One Dimension", Academic, New York, 1966.
\bibitem{mathieu3} S. S. Gubser and A. Hashimoto, Comm. Math. Phys.
203 (1999) 325, hep-th/9805140; M. Cveti\v{c}, H. L\"{u}, 
C. N. Pope and T. A. Tran, Phys. Rev. {\bf D 59} (1999) 126002, 
hep-th/9901002.
\bibitem{mathieu4} E. L. Ince, Proc. Roy. Soc. Edin. {\bf 46} (1925) 20;
S. Goldstein, Camb. Phil. Soc. Trans. {\bf 23} (1927) 303. 
\bibitem{mathieu5} ``Tables Relating to Mathieu Functions", National
Bureau of Standards, Columbia Univ. Press, New York, 1951. {\bf
See Eq. (2.35) in page XVIII of this reference for the asymptotic 
expansion, Eq. (\ref{eq:incegoldstein}) of this paper}.
\bibitem{jpa} E. M. F. Curado and M. A. Rego-Monteiro, 
J. Phys. {\bf A 34} (2001) 3253.   
\bibitem{comhugo} E. M. F. Curado, M. A. Rego-Monteiro and 
H. N. Nazareno, Phys. Rev. {\bf A 64} (2001) 12105; hep-th/0012244.
\bibitem{campos} M. A. Rego-Monteiro and E. M. F. Curado, Int. J.
Mod. Phys. {\bf A 17} (2002) 661.
\bibitem{dimakis1} A. Dimakis and F. Muller-Hoissen, Phys. Let.  
{\bf B 295} (1992) 242.
\bibitem{varyingc} J.-P. Uzan, 
Rev. of Mod. Phys {\bf 75} (2003) 403.


\end{thebibliography}
\end{document}